\documentclass[aip,reprint,superscriptaddress, showpacs, floatfix]{revtex4-2}

\usepackage{graphicx}
\usepackage{dcolumn}
\usepackage{bm}
\usepackage{xcolor,colortbl,color}
\usepackage{amsmath,amsthm}
\usepackage{hyperref}
\usepackage{xurl}
\usepackage[caption=false]{subfig}
\usepackage{floatrow}
\usepackage{braket}
\usepackage{bbold}
\usepackage{comment}
\usepackage{siunitx}
\usepackage{soul}
\usepackage{algorithm}
\usepackage[noend]{algpseudocode}
\newtheorem{proposition}{Proposition}
\graphicspath{{figures/}}
\definecolor{black!15}{gray}{0.9}
\definecolor{black!30}{gray}{0.7}
\definecolor{black!50}{gray}{0.1}

\DeclareMathOperator{\Tr}{Tr}

\begin{document}

\title{Robustness of Dynamic Quantum Control: Differential Sensitivity Bounds}

\author{S.~P.~O'Neil}
\email{seanonei@usc.edu}
\affiliation{Department of Electrical and Computer Engineering, University of Southern California, Los Angeles, CA 90089, USA}

\author{C.~A.~Weidner}
\email{c.weidner@bristol.ac.uk}
\affiliation{Quantum Engineering Technology Laboratories, H.\ H.\ Wills Physics Laboratory and Department of Electrical and Electronic Engineering, University of Bristol, Bristol BS8 1FD, UK}

\author{E.~A.~Jonckheere}
\email{jonckhee@usc.edu}
\affiliation{Department of Electrical and Computer Engineering, University of Southern California, Los Angeles, CA 90007, USA}

\author{F.~C.~Langbein}
\email{frank@langbein.org}
\affiliation{School of Computer Science and Informatics, Cardiff University, Cardiff, CF24 4AG, UK}

\author{S.~G.~Schirmer}
\email{sgs29@swan.ac.uk, lw1660@gmail.com}
\affiliation{Faculty of Science and Engineering, Physics, Swansea University, Swansea, SA2 8PP, UK}

\begin{abstract}
Dynamic control via optimized, piecewise-constant pulses is a common paradigm for open-loop control to implement quantum gates.  While numerous methods exist for the synthesis of such controls, there are many open questions regarding the robustness of the resulting control schemes in the presence of model uncertainty; unlike in classical control, there are generally no analytical guarantees on the control performance with respect to inexact modeling of the system.   In this paper a new robustness measure based on the differential sensitivity of the gate fidelity error to parametric (structured) uncertainties is introduced, and bounds on the differential sensitivity to parametric uncertainties are used to establish performance guarantees for optimal controllers for a variety of quantum gate types, system sizes, and control implementations.  Specifically, it is shown how a maximum allowable perturbation over a set of Hamiltonian uncertainties that guarantees a given fidelity error, can be reliably computed.  This measure of robustness is inversely proportional to the upper bound on the differential sensitivity of the fidelity error evaluated under nominal operating conditions. Finally, the results show that the nominal fidelity error and differential sensitivity upper bound are positively correlated across a wide range of problems and control implementations, suggesting that in the high-fidelity control regime, rather than there being a trade-off between fidelity and robustness, higher nominal gate fidelities are positively correlated with increased robustness of the controls in the presence of parametric uncertainties.
\end{abstract}

\maketitle

\section{Introduction}

A prerequisite for the widespread adoption of practical quantum devices is their ability to perform in noisy environments. Thus, one must not only consider the effective control of quantum devices but also robust quantum control, where the figure of merit is maintained in the face of noise, drift, and other deleterious effects.

Indeed, the robust control of quantum systems in the presence of noise has been considered for a variety of technologically-relevant systems, among them quantum computational architectures based on trapped ions~\cite{Hensinger_2022}; spin~\cite{Barnes_2015} and superconducting qubits~\cite{biercuk_2021, Werninghaus_2021}; atom-interferometric inertial sensors~\cite{biercuk_2023, Anderson_2017}; and linear quantum optics~\cite{Dong_2020}, among many others. There are a number of methods for direct determination and optimization of robust quantum controllers, among them composite pulses~\cite{wimperis_1994, Wilhelm_2009}, pulse shaping~\cite{Guerin_2013}, geometric control~\cite{Xue_2023}, minimization of the quantum Fisher information~\cite{Poggi_2023}, hybrid gradient-descent/simulated annealing~\cite{Mahesh_2022}, adapted Krotov for disorder-dressed evolution~\cite{Gneiting_2023}, and algorithms based on Pareto optimization~\cite{Bhutoria_2022}.

Beyond generating potential controllers, analysis methods for determining the robustness of controllers post-optimization are important. Robustness measures enable systematic comparison of controllers and facilitate the selection of those most suitable. One such measure is the Robustness Infidelity Measure (RIM)~\cite{RIM}, the lowest order of which reduces to the average fidelity error of a controller estimated by sampling over perturbations, or through propagation of uncertainties~\cite{motzoi_2022}. Robustness measures for quantum controllers have also been developed based on the control-theoretic measures of differential and logarithmic sensitivity~\cite{LCSS, Shermer_2023}, but to date, these methods have been limited to static control, or so-called \emph{energy landscape control}~\cite{Schirmer_2018}, where the control fields are time-independent and the optimal system evolves from a predetermined initial state to a desired final state.  While static control is a promising novel paradigm for certain applications, it has limitations, and the majority of control schemes proposed to-date for quantum systems are based on dynamic control. Dynamic control is usually formulated in terms of piecewise-constant control amplitudes~\cite{Khaneja_2005, Krotov}, found via optimization with respect to a basis~\cite{CRAB, GROUP}, or one of the robust controller optimization methods listed above; a survey of optimization methods can be found in~\onlinecite{Koch2022}. The robustness of dynamic control designs is typically assessed via Monte-Carlo sampling, which can be computationally expensive~\cite{motzoi_2022, Hensinger_2022}, especially for dynamic controls with similarly dynamic perturbations.

In an effort to produce analytic methods for determining the robustness of dynamic controllers, the notion of differential sensitivity has recently been extended to dynamic control with time-dependent, piecewise-constant controls~\cite{sensitivity_bounds}.  The focus of this paper is on the application of these techniques to quantum gate optimization problems with the aim to establish performance guarantees and understand trade-offs (or lack thereof) between performance and robustness, focusing on a set of benchmark problems originally laid out in Ref.~\onlinecite{Machnes_2011}.  In the years since the publication of the original problems, work has continued on the control of qubit registers for quantum device optimization~\cite{dAmico} and spintronic systems~\cite{Zinner2016, Zinner2020}.  In addition, spin chains~\cite{Shor2016} and general spin networks~\cite{Terhal2015} have been proposed as a means of performing universal quantum computation.  Nonetheless, the problems originally formulated as benchmark problems for optimization algorithms, serve as a varied and useful testbed for our methods.   Although this work specifically considers gate implementation for closed quantum systems, the methods considered here can be extended to open quantum systems subject to decoherence and dissipation in a straightforward manner.  Algorithms and methods to control such systems can be found in numerous works~\cite{Koch_2016, Kosloff_2022, Bhutoria_2022, Gneiting_2023}.  Related recent work has also explored state transfer in spin chains coupled to quantum baths~\cite{Wang2021}.

The main contribution of this work is to demonstrate the efficacy of differential sensitivity-based methods~\cite{sensitivity_bounds} as a tool to analyze the robustness of the quantum gate fidelity realized by piecewise-constant controls to parametric uncertainty.  This technique provides a valuable, analytically-based post-selector for dynamic quantum controllers, complementing more common stochastic methods~\cite{RIM,Koswara_2021}.  We also demonstrate how the system control time and time resolution of the controls affect both gate fidelity and robustness in linear qubit registers with Ising or Heisenberg-type coupling. 

The manuscript is organized as follows: Section~\ref{sec:desc} describes the problems considered, as well as the target objectives. Section~\ref{sec:diff} presents the expression for the differential sensitivity, summarizes the bounds based on this measure, and the largest perturbation allowable to guarantee a given performance requirement. Results are presented and discussed in Section~\ref{sec:results}. The relationship between differential and log-sensitivity and fundamental limitations in classical control is discussed in Section~\ref{sec: fundamental_limitations}.  Conclusions and future work are summarized in Section~\ref{sec:conc}.

\section{Systems and Control Objectives}\label{sec:desc}

\subsection{System Dynamics}

We consider the problem of maximizing the fidelity of gate operations on a linear register of qubits. The evolution of the system is governed by a control-dependent unitary propagator $U(t)$, and the target gate is given by a unitary operator $U_f$. In this work we do not consider the effects of finite qubit measurement times and assume a fixed gate operation time $t_f$, as is common in the literature. We restrict our attention to a network of $Q$ qubits with an underlying Hilbert space of dimension $N = 2^Q$. The control objective is the maximization of the normalized gate fidelity 
\begin{equation}
    \mathcal{F} = \frac{1}{N} \left| \Tr\left[ U_f^{\dagger} U(t_f) \right] \right|.
\end{equation}

In the absence of external control fields, the evolution of $U(t)$ is governed by the nominal drift Hamiltonian $H_0 \in \mathbb{C}^{N \times N}$. To modify the evolution, we employ $M$ control fields through a set of $M$ interaction Hamiltonian matrices $H_m$ for $1 \leq m \leq M$. Finally, we divide the time interval from the initial time $t_0$ to the gate operation time $t_f$ into $\kappa$ uniform time intervals of length $\Delta = t_f/ \kappa$. With the initial time as $t_0 = 0$ and final time as $t_f = \kappa \Delta$, the intermediate time steps are $t_k = k\Delta$, $0\leq k < \kappa$.

Within each interval $[t_{k},t_{k+1})$, we restrict the $M$ control fields to take constant values. Then the control pulses are described as $f_{m}^{(k)} \in \mathbb{R}$, where $m$ denotes the control channel associated with $H_m$ and $k$ the time interval starting at $t_k$. The total Hamiltonian during time interval $[t_{k},t_{k+1})$ is
\begin{equation}
    H^{(k)} = H_0 + \sum\limits_{m = 1}^M H_m f_{m}^{(k)}.
\end{equation} 
Let $\chi^{(k)}(t)$ be a constant pulse of unit magnitude that is only non-zero for the interval $[t_k,t_{k+1})$. Then, the total Hamiltonian can be written as the sum of $H^{(k)}$ over time 
\begin{equation}
    H(t) = \sum_{k = 0}^{\kappa-1} H^{(k)} \chi^{(k)}(t) 
         = H_0 + \sum_{k=0}^{\kappa-1} \sum_{m=1}^M H_m f_{m}^{(k)} \chi^{(k)}(t). 
\end{equation}

For closed systems, the dynamics are governed by the time-dependent Schr\"odinger equation 
\begin{equation}\label{eq: schrodinger_gate}
    \dot{U}(t) = -\tfrac{i}{\hbar} H(t)U(t),\quad U(0) = U_0,
\end{equation}
whose solution at the gate operation time $t_f$ is  
\begin{align}
\begin{split}
    U(t_f) &= \Phi^{(\kappa,0)}U_0 
            = \left[\prod_{k=0}^{\kappa-1} \Phi^{(k+1,k)}\right]U_0 \\
           &= \Phi^{(\kappa,\kappa-1)}\Phi^{(\kappa-1,\kappa-2)} \cdots \Phi^{(1,0)} U_0.
\end{split}
\end{align}
Here $\prod_{k=0}^{\kappa-1}$ indicates a time-ordered product where
\begin{equation}\label{eq: nominal_phi_matrix}
  \Phi^{(k+1,k)}  = \exp \left[ -\tfrac{i}{\hbar} H^{(k)}\Delta \right]
\end{equation}
is the solution to the Schr\"odinger equation on the interval $[t_{k},t_{k+1})$, and $\Phi^{(\kappa,0)}$ is the concatenation of the $\kappa$ time-ordered state transition matrices. Without loss of generality, taking the initial unitary $U_0$ to be the identity $I_N$, the figure of merit is
\begin{equation}\label{eq: nominal_fidelity}
    \mathcal{F} = \frac{1}{N}\left| \Tr \left[ U_f^{\dagger}\Phi^{(\kappa,0)} \right] \right|,
\end{equation}
and the corresponding nominal fidelity error is $\varepsilon = 1 - \mathcal{F}$.
 
\subsection{Range of Problems}\label{ss: problems}

\begin{table*}[t]
\centering
\caption{A summary of the gate fidelity problems considered. Problems $1$ to $4$, $7$, and $9$ are set to maintain notational consistency with Ref.~\onlinecite{Machnes_2011}. Problems $5$, $6$, and $8$ are newly selected problems formulated specifically for the robustness analysis presented here. \textit{Individual Qubit Addressability} denotes a control architecture in which each qubit is subject to control by an independent external field. \textit{Simultaneous Control on All Qubits} is the same external field applied to all qubits in the linear register. \textit{Initial Qubit Control Only} refers to the application of an external field to only the first qubit in the register. See Section~\ref{ss: problems} for details.}\label{table: problems}
\begin{tabular}{|c|l|l|l|l|}
    \hline
    \textbf{Problem} & \textbf{Description} & \textbf{Target gate --- $U_f$} & \textbf{$t_f$ options} & \textbf{$\kappa$ options} \\ \hline
    $1$ & Ising ZZ $2$-Qubits -- \textit{Individual Qubit Addressability}  & Controlled NOT & $2, 3, 4$ & $40, 64, 128$ \\ \hline
    $2$ & Ising ZZ $3$-Qubits -- \textit{Individual Qubit Addressability}  & Quantum Fourier Transform & $7, 8$ & $40, 64$ \\ \hline
    $3$ & Ising ZZ $4$-Qubits -- \textit{Individual Qubit Addressability}  & Quantum Fourier Transform & $12, 15, 20$ & $40, 64$ \\ \hline
    $4$ & Ising ZZ $5$-Qubit -- \textit{Individual Qubit Addressability}  & Quantum Fourier Transform & $12, 15, 25$ & $64, 128$ \\ \hline
    $5$ & Heisenberg XXX $3$-Qubits -- \textit{Individual Qubit Addressability} & Quantum Fourier Transform & $7, 8$ & $40, 64$ \\ \hline 
    $6$ & Heisenberg XXX $3$-Qubits -- \textit{Individual Qubit Addressability} & Random Unitary & $7, 8$ & $40, 64$ \\ \hline 
    $7$ & Ising ZZ $5$-Qubits -- \textit{Simultaneous Control on All Qubits} & Quantum Fourier Transform & $125,150$ & $1000$ \\ \hline
    $8$ & Heisenberg XXX $3$-Qubits -- \textit{Initial Qubit Control Only} & Quantum Fourier Transform & $10, 15$ & $32, 64$ \\ \hline 
    $9$ & Heisenberg XXX $3$-Qubits -- \textit{Initial Qubit Control Only} & Random Unitary  & $10, 15$ & $32, 64$ \\ \hline
\end{tabular}
\end{table*}

As a testbed for our robustness analysis for piecewise-constant, dynamic controllers we build on the problems presented in Ref.~\onlinecite{Machnes_2011}. These problems provide a range of system sizes and control implementations for linear qubit registers with Ising or Heisenberg-type coupling. This makes the problem selection suitable for assessing the effect of different interaction types and control settings, such as local addressability versus global control, on the robustness of the controlled system while remaining within the subset of linear qubit register architectures with nearest-neighbor coupling. 

We focus here on quantum gate implementation problems, ranging in complexity from two- to five-qubit gates.  The full list of problems considered is summarized in Table~\ref{table: problems}. The gate operation times $t_f$ and number of time steps $\kappa$ were selected to facilitate synthesis of high-fidelity controllers $\left(\mathcal{F}(t_f)>0.99 \right)$. For most problems the values in~\onlinecite{Machnes_2011} were chosen as a starting point for each problem, with gate operation times varied from the shortest times for which we can find controllers that achieve minimum performance criteria to larger $t_f$, where it is usually easier to find solutions.  Similarly, for the number of time steps $\kappa$, we typically aim to start with values just large enough to be able to find controllers that meet certain minimum performance requirements and then increase $\kappa$, which increases the dimension of the search space and therefore (generally) the set of controllers that meet minimum performance requirements.  For example, we may select minimum values for $t_f$ and $\kappa$ such that our optimization procedure yields controllers that achieve fidelities $\ge 0.99$ with at least some controllers achieving fidelities $>0.9999$. 

The starting point for all systems is a drift Hamiltonian of the form
\begin{align}\label{eq:H0-full}
    H_0  = & \frac{\hbar J}{2} \sum_{\ell = 1}^{Q-1} \left( 
             \alpha \sigma_x^{(\ell)} \sigma_x^{(\ell+1)}
             + \alpha \sigma_y^{(\ell)} \sigma_y^{(\ell+1)}
             + \beta  \sigma_z^{(\ell)} \sigma_z^{(\ell+1)} \nonumber \right) \\
           & + \frac{\hbar}{2} \sum_{\ell=1}^Q \omega_\ell\sigma_z^{(\ell)},
\end{align}
where $Q$ is the number of qubits, $\sigma_{\set{x,y,z}}^{(\ell)}$ denotes the $Q-$fold tensor product of $I_2$ with $\sigma_{\set{x,y,z}}$ in the $\ell$-th position, and $\{\alpha,\beta\} \in \{0,1\}$ differentiate the drift Hamiltonian for distinct problems as described below. The matrices $\sigma_{\set{x,y,z}}$ are the $2 \times 2$ Pauli matrices, and $I_2$ is the identity. The second term in the drift Hamiltonian corresponds to local on-site potentials \textcolor{black}{with transition frequencies $\omega_\ell$}, while the first term represents couplings between adjacent qubits, which are assumed to be uniform and fixed. In the following, we choose energy in units of $\hbar J$ and time in units of $J^{-1}$,  drop the factors of $\hbar J$. 
\textcolor{black}{If we assume that the transition frequencies are sufficiently well-separated to prevent off-resonant excitation and the control pulses vary much more slowly than the transition frequencies, then we can transform into a rotating frame and employ the rotating wave approximation in which the counter-rotating terms cancel over the cycle of a control pulse~\cite{schirmer_2006_hamiltonian_engineering} to eliminate 
the local potentials,}
in which case the drift Hamiltonian simplifies to
\begin{equation}\label{eq: XXX}
    H_0  = \frac{1}{2} \sum_{\ell = 1}^{Q-1} \left( 
             \alpha \sigma_x^{(\ell)} \sigma_x^{(\ell+1)}
           + \alpha \sigma_y^{(\ell)} \sigma_y^{(\ell+1)}
           + \beta  \sigma_z^{(\ell)} \sigma_z^{(\ell+1)} \right).
\end{equation}
For the \textit{Individual Qubit Addressability} subset of problems, we assume all qubits are individually addressable and that we can perform local $x$ and $y$ rotations, leading to $M=2Q$ control Hamiltonians
\begin{equation}\label{eq: Hc-full}
    H_{2m-1} = \frac{1}{2} \sigma_x^{(m)}, \quad 
    H_{2m}   = \frac{1}{2} \sigma_y^{(m)}, \quad 1 \leq m \leq M.
\end{equation}

For Problems~$1$ to $4$ we assume full local control given by Eq.~\eqref{eq: Hc-full} and \emph{Ising coupling} between adjacent qubits, setting $\alpha=0$ and $\beta=1$ in Eq.~\eqref{eq: XXX}. For Problem~$1$, the target unitary is a CNOT gate. For Problems~$2$, $3$, and $4$, the Quantum Fourier Transform (QFT) gate was chosen as it is defined for an arbitrary number of qubits and plays a key role in many quantum algorithms. The matrix elements of the QFT gate are
\begin{equation}\label{eq:QFT}
    \mathrm{QFT}_{(j,k)} = \frac{1}{\sqrt{N}}\omega^{jk},
\end{equation}
where $\omega = \exp{(2\pi i/N)}$ such that $\omega^{N} = 1$ and $\omega$ is raised to the $jk$-th power in the $(j,k)$-th entry of the matrix representing the gate, where $j,k = 0, \ldots, N-1$.

To assess whether the interaction type affects the sensitivity behavior we consider a qubit register with Heisenberg coupling, for which we set $\alpha=\beta=1$ in Eq.~\eqref{eq: XXX}, again assuming full local control given by Eq.~\eqref{eq: Hc-full}. Specifically, we compare the implementation of a three-qubit QFT gate for a Heisenberg-coupled system (Problem~$5$) with its Ising equivalent (Problem~$2$). To eliminate target gate dependence and amplify the control difficulty, a random unitary operator $U_f \in \mathbb{U}(8)$, with elements distributed by Haar measure on $\mathbb{U}(8)$~\cite{Mezzadri_2007,Bredon1993,Jonckheere1997}, is chosen as a target gate for Problem~$6$.

Another aspect that could plausibly influence the robustness of the optimized controllers is the type of control. Individual spin addressability generally affords the most control, but this is not always required. For example, it often suffices to have control over a single qubit at one end of the register as the effect of the controls are propagated along the chain by Heisenberg (but not Ising-type) coupling~\cite{PhysRevA.78.062339}. To assess the effect of such restricted control, we include the implementation of a three-qubit QFT and random unitary gate for a Heisenberg-coupled register with control of the first qubit only (Problems~$8$, $9$). We designate this control implementation as \textit{Initial Qubit Control Only} in Table~\ref{table: problems}. For these problems, the control Hamiltonians reduce to $H_{1} = \tfrac{1}{2} \sigma_{x}^{(1)}$ and $H_{2} = \tfrac{1}{2} \sigma_{y}^{(1)}$. For Problem~$8$ the target unitary is a QFT, while for Problem~$9$ the target is the same random unitary used for Problem~$6$.

Selective addressing of qubits can be achieved in certain control settings such as laser control of trapped ions, atoms or NV centers, where a laser can be focused on individual qubits, or quantum registers where qubits are controlled by surface gate electrodes. However, direct selective addressing is often not possible. For example, in the microwave regime, focusing microwaves on individual qubits is challenging, and in molecular systems and applications involving nuclear magnetic resonance (NMR) or electron spin resonance (ESR), magnetic fields cannot be focused directly on a single nuclear or electron spin. In this case selectivity is typically achieved by frequency-selective addressing, either by taking advantage of existing chemical shifts or by applying electric or magnetic field gradients to ensure that different qubits have different resonance frequencies. However, frequency-selective addressing has drawbacks; in particular it requires that the control amplitudes must be modulated on time-scales that are slow compared to the frequency difference between qubits to minimize off-resonant excitation~\cite{Nigmatullin_2009}. In these cases it is often preferable to use a global control model. To cover this case, in Problem~$7$ we consider a linear qubit register with Ising coupling between adjacent qubits as before but adding a position-dependent Stark or Zeeman shift, which leads to a drift Hamiltonian where the local onsite potentials do not vanish:
\begin{equation}\label{eq:ZZ_Stark}
    H_0 = \frac{1}{2} \sum_{\ell = 1}^{Q-1} \sigma_z^{(\ell)} \sigma_z^{(\ell+1)} - \frac{1}{2} \sum_{\ell=1}^Q (\ell + 2) \sigma_z^{(\ell)}.
\end{equation}
Since the controls are now simultaneously acting on all qubits, the control Hamiltonian matrices are 
\begin{equation}\label{eq: Hc-global}
    H_1 = \frac{1}{2} \sum_{\ell = 1}^Q \sigma_x^{(\ell)}, \quad
    H_2 = \frac{1}{2} \sum_{\ell = 1}^Q\sigma_y^{(\ell)},
\end{equation}
and we denote this implementation as \textit{Simultaneous Control on All Qubits} in Table~\ref{table: problems}. We choose $Q=5$ with the target gate being a five-qubit QFT gate as in Problem~$4$.

\subsection{Controller Synthesis}

The focus of this paper is the analysis of the robustness of controllers. The techniques presented are independent of the synthesis method used and can be applied to controllers obtained from arbitrary algorithms. Here, we use standard algorithms similar to those used in Ref.~\onlinecite{Floether_2012} to generate the controllers. Unless otherwise stated, the controllers analyzed in this paper were generated by unconstrained optimization in MATLAB using the \texttt{fminunc} function, with the goal of producing an optimal sequence of control fields $\set{f_{m}^{(k)}}_{k = 0}^{\kappa-1}$ that minimize $\varepsilon$ with $\kappa$ time steps. The initial condition for the optimization problem was an $M \times \kappa$ array of initial control fields for each control channel and time interval $[t_k,t_{k+1})$. The initial values were either drawn randomly from a uniform distribution or a standard normal distribution, and for most problems one initial condition with all fields set to $0$ was also included.

Controllers were generated using both the \texttt{quasi-newton} and \texttt{trust-region} options for the optimization algorithm. Unless explicitly noted, the controllers found for each algorithm are not differentiated, as the focus of this paper is not on algorithm comparison, although we note that the specific control algorithm choices may bias statements about the overall controller properties. Contingent on the problem, we choose gate operation times between $t_f = 2$ and $t_f = 150$ in units of $1/J$. The number of time steps was varied between $\kappa = 32$ and $\kappa = 1000$. As mentioned above, these parameters were chosen to ensure synthesis of controllers with a fidelity error less than $10^{-2}$. After filtering out controllers with $\epsilon \geq 10^{-2}$, this process provides between $22$ and $100$ controllers for each combination of problem, $t_f$, and $\kappa$.

\section{Robust Performance in the Presence of Uncertainty}\label{sec:diff}

Predictions of any model of an experimental system are subject to uncertainty in practice. Understanding the sensitivity of the control performance with regard to various uncertainties is of utmost importance. This can be done numerically by Monte Carlo simulations, randomly varying model parameters and performing statistical analysis. However, this is computationally intensive and may not provide deeper insight into which uncertainties are most critical in terms of robustness.  Here, we consider an alternative approach that quantifies the effect of \emph{structured} perturbations to the Hamiltonian on the fidelity error in terms of the differential sensitivity of the fidelity error~\cite{sensitivity_bounds}.  We establish the uncertainty model for the sensitivity analysis, present the differential sensitivity of the fidelity error for dynamic controls, and discuss upper bounds on the latter that are agnostic to the exact structure of the uncertainty.

\subsection{Uncertainty Model}\label{ss:uncertainty_model}

Uncertainties in model parameters often lead to structured perturbations to the Hamiltonian. For example, uncertainty in the $J$-coupling between spins will manifest itself in a structured uncertainty to the drift Hamiltonian. The structure depends on which couplings are affected, if they are affected independently or collectively, etc. \textcolor{black}{Formally, we model structured uncertainties to the Hamiltonian $H(t)$ in the $k$th time-step as $\delta H^{(k)}_\mu$, where $\delta \in [\delta_1,\delta_2]$ is the scalar deviation of the uncertain (set of) parameter(s) from their nominal values, and 
\[ H^{(k)}_{\mu}:=\sum_{m=0}^M s_m^{(k)} \hat{H}_m \alpha_m^{(k)}
\] represents the scaled structure of the uncertainty in time step $k$. Each $\hat{H}_m$ is a Hermitian matrix, normalized with respect to the Frobenius norm such that $\|\hat{H}_m\|_{F} = 1$. Specifically, $\hat{H}_0$ models the uncertainty in the drift Hamiltonian, while the $\hat{H}_m$ for $1 \leq m \leq M$ model the uncertainty in the control Hamiltonians. The set $\set{s^{(k)}_m}$ are scalar weights, normalized so that $\mathbf{s}^{(k)}_\mu = (s^{(k)}_0,\ldots,s^{(k)}_M)$ retains 
normalization of the un-scaled structure $\sum_{m=0}^M s^{(k)}_m \hat{H}_m$.
The scalars $\alpha_m^{(k)}$ scale the structure based on the uncertainty type considered and described below.}

\textcolor{black}{We write the perturbed Hamiltonian $\tilde{H}(t)$ as a sum of the perturbed Hamiltonians $\tilde{H}^{(k)}$ for the $k$th time step:
\begin{subequations}\label{eq: perturbed_hamiltonian_1}  \begin{align}
    \tilde{H}(t) &= \sum_{k=0}^{\kappa-1} \tilde{H}^{(k)} \chi^{(k)}(t), \\ 
    \tilde{H}^{(k)}  &= H_0 + \sum_{m=1}^{M} H_m f_{m}^{(k)} + \delta {H}_\mu^{(k)} \\ 
    &= H^{(k)} + \delta \sum_{m=0}^M s^{(k)}_m \hat{H}_m \alpha_m^{(k)}. 
\end{align}\end{subequations}}
\textcolor{black}{For uncertainty in the drift Hamiltonian, we have $\tilde{H}_0 = H_0 + \delta s^{(k)}_0 \hat{H}_0$ so that $\alpha_0^{(k)} = 1$ for all $k$. For uncertainty in the $m$th control Hamiltonian over time step $k$ we have $\tilde{H}_m f_m^{(k)}  = (H_m + \delta s^{(k)}_m \hat{H}_m)f_m^{(k)} = H_m f_m^{(k)} + \delta s^{(k)}_m \hat{H}_m f^{(k)}_m$, so that $\alpha_m^{(k)} = f_m^{(k)}$.} 

This formulation is very general. For example, we could model an uncertainty in a single $J$-coupling between qubits $1$ and $2$, resulting in a deviation from the nominal value $J$, by setting ${H}_\mu = \frac{1}{2} \sigma_z^{(1)} \sigma_z^{(2)}$ with $\delta = J_{12}-J$. 

\subsection{Fidelity Error and Differential Sensitivity}\label{ss:  differential_sensitivity_derivation}

We ultimately desire performance criteria that ensure the \emph{perturbed fidelity error} 
\begin{equation}\label{eq: error(delta)}
   \tilde{\varepsilon}_{\mu}(\delta) = 1 - \frac{1}{N} \left| \Tr\left[U_f^{\dagger} \tilde{\Phi}^{(\kappa,0)} (\delta) \right] \right|
\end{equation}
as a function of the uncertainty size $\delta$ remains below a certain acceptable threshold $\epsilon$ for perturbations $\delta$ smaller than a certain (critical) value $\bar{\delta}$. Here we make explicit the dependence of the perturbed propagator on the uncertainty strength $\delta$ as   
\begin{equation}
    \tilde{\Phi}^{(\kappa,0)}(\delta) = \prod_{k = 0}^{\kappa-1} \tilde{\Phi}^{(k+1,k)}(\delta)  = \prod_{k=0}^{\kappa - 1}  \exp \left[-i\tilde{H}^{(k)}\Delta \right].
\end{equation}
 
In principle this can be done by numerically evaluating Eq.~\eqref{eq: error(delta)} for a given uncertainty structure ${H}_\mu$ and a range of strengths $\delta$ to determine the $\bar{\delta}$ at which $\tilde{\varepsilon}_\mu(\bar{\delta}) \geq \epsilon$, similar to what is explored in Ref.~\onlinecite{RIM}. A drawback of this approach is the computational cost involved, potentially requiring many fidelity evaluations, and the difficulty establishing analytic results.

Here we use the differential sensitivity of the perturbed fidelity error to establish performance guarantees. \textcolor{black}{If the uncertainty structure is constant over time, i.e., $\mathbf{s}^{(k)}_\mu = \mathbf{s}_\mu$ for all $k$, then Ref.~\onlinecite{sensitivity_bounds} shows that we can write}
\begin{equation}\label{eq: Z_Gamma}
\zeta_{\mu} :=
\left. \frac{\partial \tilde{\varepsilon}_\mu(\delta)}{\partial \delta} \right|_{\delta = 0} 
= \sum_{k=0}^{\kappa-1} \sum_{m=0}^M Z_{m}^{(k)} s_m
=: \mathbf{\Gamma}\mathbf{s_\mu},
\end{equation}
where 
\begin{equation}\label{eq: Z_km}
    Z_m^{(k)} = \Re \left\{ -\frac{e^{-i\phi}}{N} \Tr\left[\Phi^{(k,0)}U_f^\dagger \Phi^{(\kappa,k-1)} \mathbf{X}^{(k)}_m \right] \right\},
\end{equation}
\begin{equation}\label{eq: X_km}
  \mathbf{X}^{(k)}_m = -i
  \int_{t_{k}}^{t_{k+1}} e^{-iH^{(k)}(t_{k+1} -\tau)} \hat{H}_m \alpha_{m}^{(k)} e^{-iH^{k} (\tau-t_{k})}d \tau,
\end{equation}
and $\phi = \angle \Tr\left[ U_f^{\dagger} \Phi^{(\kappa,0)} \right]$.

Eq.~\eqref{eq: Z_Gamma} facilitates two interpretations of the differential sensitivity. If the uncertainty is constant for the entire evolution then $\mathbf{s}_\mu$ is a constant vector and $\mathbf{\Gamma}$ can be viewed as a matrix operator that accepts as input a static uncertainty structure, $\mathbf{s_\mu}$, and provides as output the sensitivity in that direction. In this case, a \emph{static-uncertainty maximum bound} on the differential sensitivity at $\delta = 0$ can be derived~\cite{sensitivity_bounds}.

If the uncertainty is constant for each time step but varies between time steps then \textcolor{black}{we have the more general case $H_\mu^{(k)} = \sum_{m=0}^{M} s_m^{(k)}\hat{H}_m \alpha_m^{(k)}$} with corresponding $\mathbf{s}_\mu^{(k)}$ for each time step $k$. The differential sensitivity can thus be written as
\begin{multline}\label{eq: Z_time_change}
    \zeta_{\set{\mu}} = 
    \sum_{k=0}^{\kappa-1} \left(\sum_{m=0}^{M} Z_{m}^{(k)} {s}_m^{(k)} \right) 
    = \sum_{k=0}^{\kappa-1} \mathbf{Z}^{(k)} \mathbf{s}_{\mu}^{(k)} 
\end{multline}
where $\zeta_{\{\mu\}}$ indicates that $H_\mu$ is not fixed but given by the sequence $\{H_\mu^{(k)}: k=0,\ldots, \kappa-1 \}$. In this case an alternative \emph{variable-uncertainty \textcolor{black}{(\textit{vu})} maximum bound}, $\mathcal{B_{\text{vu}}}$, on the differential sensitivity at $\delta = 0$ can be derived~\cite{sensitivity_bounds}:

\begin{proposition}\label{prop: prop_1} 
If the uncertainty can vary from one time step to the next, and is described by the sequence of structure matrices $\set{\hat{H}_{\mu}^{(k)}}_{k=0}^{\kappa-1}$, the variable-uncertainty upper bound on $\|\zeta_{\set{\mu}} \|$ is given 
$\mathcal{B_{\text{vu}}} = \| \set{\bar{\varsigma}^{(k)}} \|_{\ell^1}$,
the $\ell^1$ norm of the sequence $\set{\bar{\varsigma}^{(k)}}$, with $\bar{\varsigma}^{(k)} = \|\mathbf{Z}^{(k)} \|_{2}$.
The sequence $\set{\bar{\mathbf{s}}_{\mu}^{(k)}}$ that achieves the bound $\mathcal{B_{\text{vu}}}$ is $\bar{\mathbf{s}}_{\mu}^{(k)} = {\mathbf{Z}^{(k)}}^T/\bar{\varsigma}^{(k)}$ for $0 \leq k < \kappa$.
\end{proposition}

The basic idea behind this proposition is that for each time step we want the size of the differential sensitivity in the direction that maximizes the sensitivity, which is given by the magnitude (or usual $2$-norm) of the vector $\mathbf{Z}^{(k)}$.  This yields a sequence of positive numbers over $\kappa$ time steps, and the $\ell_1$ norm of this sequence provides an upper bound $\mathcal{B}_\text{vu}$ for the differential sensitivity.  Using $\mathcal{B}_\text{vu}$ and the associated sequence of maximum sensitivity directions $\set{\bar{\mathbf{s}}_\mu^{(k)}}$ allows a directed search to find the maximum perturbation strength $\bar{\delta}$ that guarantees the performance criterion $\tilde{\varepsilon}_\mu (\delta) < \epsilon$ is satisfied in any direction provided $\delta \le \bar{\delta}$~\cite{sensitivity_bounds}.

Specifically, we quantize the uncertainty size $\delta$ into uniform steps of a given magnitude $\mathtt{d}$. \textcolor{black}{We choose the step size $\mathtt{d}$ small enough so that evaluation of the fidelity for a perturbation of size $\mathtt{d}$ in the worst-case direction $\set{\mathbf{s}^{(k)}_\mu}$ remains small. Specifically, we define small as the relative error between the nominal fidelity error $\varepsilon$ and the perturbed fidelity error as $\tilde{\varepsilon}_\mu(\mathtt{d}) < 1/10 $}. We initialize the sequence of maximum sensitivity directions at $\delta = 0$ as $\set{\mathbf{s}_\mu^{(k)}(0)} = \set{\bar{\mathbf{s}}_\mu^{(k)} }$ following Proposition~\ref{prop: prop_1}.  We then update the perturbed Hamiltonian in this direction at the strength $\mathtt{d}$ as 
\[ 
  \tilde{H}^{(k)}(\delta_1) = H_0 + \sum_{m=1}^{M} H_m f_m^{(k)} + \mathtt{d} \sum_{m=0}^M \hat{H}_m f_m^{(k)}\alpha_m^{(k)}s_m^{(k)}(0)
\]
for each $k$. At each successive step of $n\mathtt{d} = \delta_n$ we compute the directions of maximum sensitivity of the fidelity error as $\set{\mathbf{s}^{(k)}_\mu(n)}$ for each $k$ as per Proposition~\ref{prop: prop_1} where $\mathbf{s}_{\mu}^{(k)}(n) = \left[s_0^{(k)}(n),s_1^{(k)}(n),\hdots,s_{M}^{(k)}(n)\right]^T$ for each $k$ generated by the Hamiltonian $\tilde{H}^{(k)}(\delta_n)$ at strength $\delta_n$. We then continue to iterate on this process until finding the $\delta_n$ such that $\tilde{\varepsilon}_\mu (\delta_n) > \epsilon$. The procedure is summarized in Algorithm~\ref{algorithm}.

\begin{algorithm}
\begin{algorithmic}[1]
\caption{Compute largest $\bar{\delta}$ such that $\tilde{\varepsilon}_\mu(\bar{\delta}) < \epsilon$}
\State{$n = 1$}\label{algorithm}
\State{Initialize Worst-Case Perturbed Hamiltonian:\\
\mbox{$\tilde{H}^{(k)}(\delta_{1}) = H_0 + \sum\limits_{m=1}^M H_m f_{m}^{(k)} + \mathtt{d} \sum\limits_{m = 0}^{M} \alpha_{m}^{(k)} \hat{H}_{m} s_m^{(k)}(0) \ \forall k$}}
\State{Evaluate $\tilde{\varepsilon}_\mu(\delta_1)$}
\While{$\epsilon - \tilde{\varepsilon}_\mu(\delta_n)>0$}
\State{$n = n+1$, $\delta_n = n \mathtt{d}$}
\State{Compute $s_\mu^{(k)}(n-1) \ \forall k$ from Proposition~\ref{prop: prop_1}}
\State{$\tilde{H}^{(k)}(\delta_n) = \tilde{H}^{(k)}(\delta_{n-1}) + \mathtt{d} \sum\limits_{m=0}^{M} \alpha_{m}^{(k)} \hat{H}_m s_m^{(k)}(n-1) \ \forall k$}
\State{Evaluate $\tilde{\varepsilon}_\mu(\delta_n)$}
\EndWhile
\State{Set $\bar{n} = n - 1$, $\bar{\delta} = \bar{n}\mathtt{d}$}
\end{algorithmic}
\end{algorithm}

\section{Application to Robustness of Quantum Gate Implementation}\label{sec:results}

\begin{table}
\caption{Results of hypothesis tests for correlation between $\mathcal{B}_\text{vu}$ and the minimum performance-violating perturbation $\bar{\delta}$ based on the Pearson correlation coefficient $r$. $\#$ is the number of samples included in the test (i.e., the number of controllers found with $\mathcal{F}>0.99$), $Z_r$ is the normalized test statistic, and $p_r$ is the $p$-value of the test. Note the strong negative trend between $\mathcal{B}_\text{vu}$ and $\bar{\delta}$. The ten cases that fail to meet the significance level of $p<0.05$ are shaded.}   \label{table: B_vu correlation}
\begin{tabular}{|c|c|c|c|c|c|c|}
    \hline
        \textbf{Problem} & \textbf{$t_f$} & \textbf{$\kappa$} & \textbf{$\#$} & \textbf{$r$} & \textbf{$Z_r$} & \textbf{$p_r$} \\ \hline
        1 & 2 & 40 & 99 & -0.210 & -2.119 & 0.018 \\ \hline
        1 & 2 & 64 & 99 & -0.327 & -3.405 & 0.000 \\ \hline
        1 & 2 & 128 & 100 & -0.376 & -4.014 & 0.000 \\ \hline
        1 & 3 & 40 & 99 & -0.282 & -2.889 & 0.002 \\ \hline
        1 & 3 & 64 & 100 & -0.281 & -2.895 & 0.002 \\ \hline
        1 & 3 & 128 & 100 & -0.180 & -1.807 & 0.037 \\ \hline
        1 & 4 & 40 & 99 & -0.258 & -2.635 & 0.005 \\ \hline
        1 & 4 & 64 & 99 & -0.556 & -6.586 & 0.000 \\ \hline
        1 & 4 & 128 & 100 & -0.265 & -2.724 & 0.004 \\ \hline
        2 & 7 & 40 & 91 & -0.444 & -4.681 & 0.000 \\ \hline
        2 & 7 & 64 & 83 & -0.309 & -2.928 & 0.002 \\ \hline
        2 & 8 & 40 & 99 & -0.660 & -8.659 & 0.000 \\ \hline
        2 & 8 & 64 & 99 & -0.483 & -5.426 & 0.000 \\ \hline
        3 & 12 & 40 & 98 & -0.551 & -6.471 & 0.000 \\ \hline
        3 & 12 & 64 & 99 & -0.500 & -5.693 & 0.000 \\ \hline
        3 & 15 & 40 & 99 & -0.406 & -4.372 & 0.000 \\ \hline
        3 & 15 & 64 & 100 & -0.437 & -4.807 & 0.000 \\ \hline
        3 & 20 & 40 & 100 & -0.322 & -3.364 & 0.001 \\ \hline
        3 & 20 & 64 & 100 & -0.477 & -5.372 & 0.000 \\ \hline
        \rowcolor[gray]{0.8}4 & 12 & 64 & 47 & -0.060 & -0.406 & 0.343 \\ \hline
        \rowcolor[gray]{0.8}4 & 12 & 128 & 34 & -0.124 & -0.708 & 0.242 \\ \hline
        4 & 15 & 64 & 91 & -0.611 & -7.290 & 0.000 \\ \hline
        4 & 15 & 128 & 91 & -0.751 & -10.725 & 0.000 \\ \hline
        4 & 25 & 64 & 96 & -0.573 & -6.784 & 0.000 \\ \hline
        4 & 25 & 128 & 100 & -0.643 & -8.315 & 0.000 \\ \hline
        5 & 7 & 40 & 100 & -0.180 & -1.810 & 0.037  \\ \hline
        \rowcolor[gray]{0.8}5 & 7 & 64 & 100 & -0.119 & -1.185 & 0.120  \\ \hline
        \rowcolor[gray]{0.8}5 & 8 & 40 & 100 & -0.052 & -0.511 & 0.305  \\ \hline
        \rowcolor[gray]{0.8}5 & 8 & 64 & 100 & -0.055 & -0.543 & 0.294  \\ \hline
        6 & 7 & 40 & 100 & -0.272 & -2.802 & 0.003  \\ \hline
        \rowcolor[gray]{0.8}6 & 7 & 64 & 100 & -0.127 & -1.266 & 0.104  \\ \hline
        \rowcolor[gray]{0.8}6 & 8 & 40 & 100 & -0.022 & -0.214 & 0.415  \\ \hline
        \rowcolor[gray]{0.8}6 & 8 & 64 & 100 & -0.014 & -0.134 & 0.447  \\ \hline
        7 & 125 & 1000 & 100 & -0.711 & -10.008 & 0.000 \\ \hline
        7 & 150 & 1000 & 100 & -0.680 & -9.186 & 0.000 \\ \hline
        8 & 10 & 32 & 33 & -0.613 & -4.323 & 0.000  \\ \hline
        8 & 10 & 64 & 97 & -0.712 & -9.880 & 0.000  \\ \hline
        8 & 15 & 32 & 100 & -0.484 & -5.473 & 0.000  \\ \hline
        \rowcolor[gray]{0.8}8 & 15 & 64 & 100 & -0.024 & -0.235 & 0.407  \\ \hline
        9 & 10 & 32 & 41 & -0.541 & -4.015 & 0.000 \\ \hline
        9 & 10 & 64 & 99 & -0.657 & -8.581 & 0.000 \\ \hline
        9 & 15 & 32 & 100 & -0.345 & -3.638 & 0.000 \\ \hline
        \rowcolor[gray]{0.8}9 & 15 & 64 & 100 & -0.065 & -0.643 & 0.261 \\ \hline
    \end{tabular}
\end{table}

Following the procedure in Algorithm~\ref{algorithm}, we can determine the minimum performance-violating perturbation for our test problems, and test the correlation between the local sensitivity at $\delta=0$, as determined by $\mathcal{B}_\text{vu}$, and the $\bar{\delta}$ that guarantees a performance threshold of $\tilde{\varepsilon}(\delta) < \epsilon$ by a one-tailed hypothesis test. We test for negative correlation indicative of a trade-off in local sensitivity around $\delta =0$ and the maximum allowable perturbation $\bar{\delta}$ that violates the performance criteria at larger perturbation strength. Using the Pearson $r$~\cite{stats_text} as the correlation measure reveals a strong negative correlation between $\mathcal{B}_\text{vu}$ and $\bar{\delta}$. Establishing the null hypothesis $\mathcal{H}_0$ as no trend between $\mathcal{B}_\text{vu}$ and $\bar{\delta}$, and $\mathcal{H}_1$ as a negative correlation between the metrics with a statistical significance level of $95\%$ ($p = 0.05)$, Table~\ref{table: B_vu correlation} shows rejection of $\mathcal{H}_0$ in favor of $\mathcal{H}_1$ for the majority of test cases. Of the $43$ test cases involving \texttt{quasi-newton}-optimized controllers, all show a negative correlation.  Ten tests fail to meet the statistical significance criteria, of which eight are Heisenberg-coupled chains while the other two are Ising chains.  While not an exact analytic relation, this suggests that the strictly local differential sensitivity bound $\mathcal{B}_\text{vu}$ is statistically correlated with robustness to larger perturbation values, agnostic of the structure or direction of the perturbation. Specifically, a higher value of $\mathcal{B}_\text{vu}$, indicating a greater potential differential sensitivity, correlates with a smaller value of the minimum performance-violating perturbation $\bar{\delta}$, indicative of a smaller margin for acceptable uncertainty.

Having established an upper bound $\mathcal{B}_\text{vu}$ on the differential sensitivity of the fidelity error and demonstrated its utility as a local measure of sensitivity and indicator of robustness, we examine the correlation between robustness quantified by $\mathcal{B}_\text{vu}$ and performance quantified by the nominal fidelity error $\varepsilon$ for the control problems with different characteristics introduced earlier.

\subsection{Fidelity Error versus Differential Sensitivity Trends}\label{ssec: error_v_sens}

\begin{figure}[t]
\subfloat[\label{fig: prob4_consolidated_scatter}]
{\includegraphics[width = 1\textwidth]{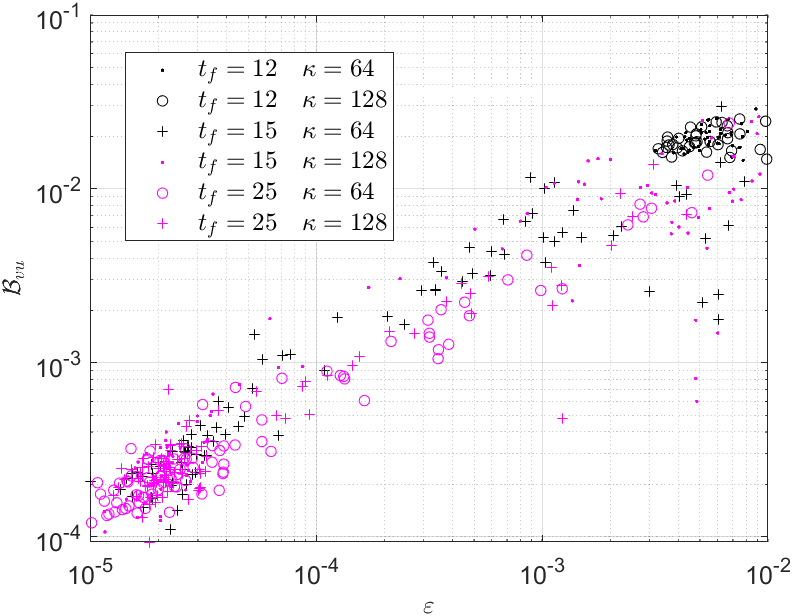}}
\hfill
\subfloat[\label{fig: prob9_consolidated_scatter}]{\includegraphics[width = 1\textwidth]{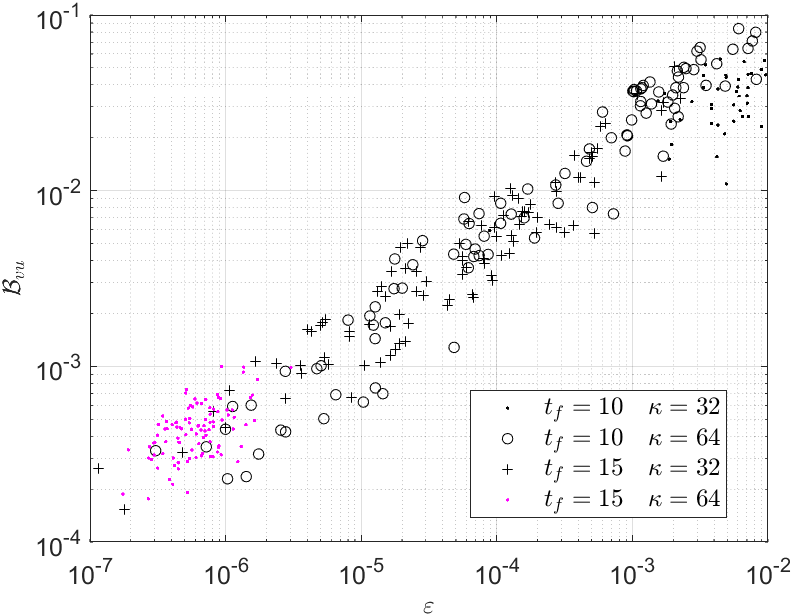}}
\caption{Plots of $\mathcal{B}_{vu}$ versus $\varepsilon$ on a log-log scale for the controllers associated with (a) Problem $4$ and (b) Problem $9$ with different gate operation times $t_f$ and time resolutions $\kappa$. These plots show a strong positive trend between the differential sensitivity bound and fidelity error for a range of gate operation times and time-resolutions.} \label{fig: scatter_prob9_loglog}  
\end{figure}

\textcolor{blue}{The upper bound on the differential sensitivity $\mathcal{B}_\text{vu}$ and the nominal fidelity error $\varepsilon$ show a strong positive correlation as indicated in Figure~\ref{fig: scatter_prob9_loglog}.  In particular, measuring the linear correlation between these metrics by the Pearson $r$ and testing for positive correlation via a one-tailed hypothesis test indicates this concordant relationship. Of the $86$ cases considered, only one does not meet the significance level of $95\%$. For the remaining $85$, the null-hypothesis $\mathcal{H}_0$ of no correlation is rejected in favor of $\mathcal{H}_1$, i.e., positive correlation between $\mathcal{B}_\text{vu}$ and $\varepsilon$. This trend is readily evident in Figs.~\ref{fig: prob4_consolidated_scatter} and \ref{fig: prob9_consolidated_scatter}, which display $\mathcal{B}_\text{vu}$ as function of $\varepsilon$ on a log-log scale for the different $(t_f, \kappa)$ cases considered for Problems~$4$ and $9$, respectively.  These plots are characteristic of the results for all problems.  In addition to the highly linear trend observed, we note a clustering of controllers in the lower-left corner of the plot in Fig.~\ref{fig: prob4_consolidated_scatter}, the space occupied by controllers with good performance (low error) and low sensitivity. Furthermore, we find that this cluster is dominated by long-time and high-$\kappa$ controllers, indicating that long control time and higher time resolution of the controls leads to more robust controllers. }

These findings are significant in that a considerable amount of work has been done on trying to simultaneously optimize performance and robustness~\cite{Kosut_2013, Koswara_2021, Koswara_2021_NJP, Khaneja_2005, Ge_2019}. Much of this work is predicated on the assumption that there is competition between performance (low fidelity error) and robustness of the controllers. The concordant relationship we observe here, however, suggests that the best performing controllers are also the most robust, at least with regard to the chosen differential sensitivity robustness measure for the problems considered. This concordance suggests that it may not be necessary to perform computationally expensive average fidelity optimization in many cases.  Furthermore, this robustness is not limited to uncertainty in the controls but extends to uncertainty in the system and interaction Hamiltonians and other structured perturbations. This positive correlation between the fidelity error and the differential sensitivity has been seen before in the context of time-invariant control of quantum state transfer~\cite{statistical_control, Schirmer_2017}. One limitation is the applicability of the differential sensitivity bound, which by its nature is mainly valid in the small perturbation regime.  For larger perturbations, it is likely that trade-offs between performance and robustness will emerge.  However, it has also been noted that broader peaks in the optimization landscape may be both more robust with regard to noise, and easier to find, especially for certain algorithms such as reinforcement learning~\cite{CDC2021-Irtaza, PhysRevResearch.5.043002}, which suggests that, at least in some cases, robust solutions with regard to noise may actually be easier to find than less robust solutions.  Less intuitively, it appears that robustness with regard to control imperfections often also implies robustness with regard to other perturbations.  An interesting question here is whether there is a maximum uncertainty for a problem where the fidelity error, sensitivity and other robustness measures such as the RIM agree, and how problem-specific such as maximum would be.

\subsection{Sensitivity and System Size}

\begin{figure*}
\subfloat[\label{fig: consolidated_scatter_1-4_0.1}]{\includegraphics[width=0.49\textwidth]{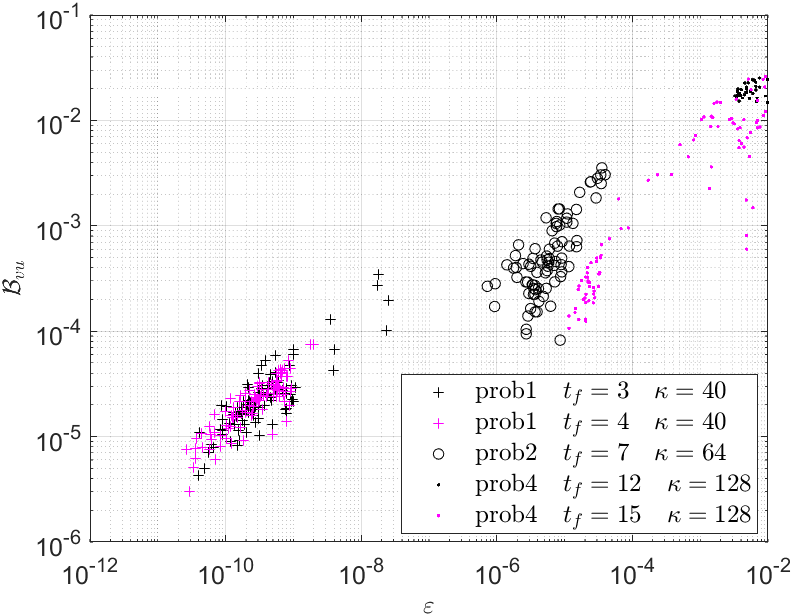}}
\hfill
\subfloat[\label{fig: consolidated_scatter_1-4_0.2}]{\includegraphics[width=0.49\textwidth]{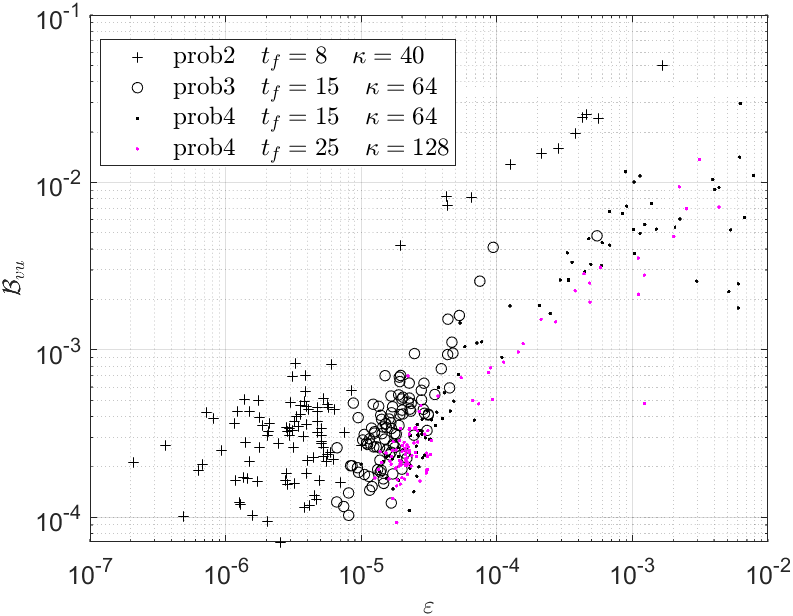}}
\caption{Plots of $\mathcal{B}_\text{vu}$ versus $\varepsilon$ on a log-log scale for controllers with similar $\Delta = t_f/\kappa$ where (a) shows $t_f/\kappa \approx 0.1$ for Problems~$1$,~$2$ and~$4$ and (b) shows $t_f/\kappa \approx 0.2$ in Problems~$2$ to~$4$. These plots show that for increasing system size and a common value of $t_f/\kappa$, robustness decreases with system size. This observation is more pronounced for the $t_f/\kappa \approx 0.1$ case in (a) than for $t_f/\kappa \approx 0.2$ case in (b).}\label{fig: consolidated_scatter_prob}
\end{figure*}

To examine the relation between the nominal fidelity error and the differential sensitivity bound as a function of system size, \textcolor{black}{we vary the number of qubits while holding other variables such as the coupling model and control implementation constant.  To this end,} we compare the results for Problems~$1$ through~$4$, which employ the same system type and control implementation (fixed Ising coupling with individual spin addressability) for system sizes increasing from two to five qubits. Fig.~\ref{fig: consolidated_scatter_prob} shows the relationship between $\mathcal{B}_\text{vu}$ and $\varepsilon$ for Problems~$1$ to~$4$ with controllers optimized for two common values of $\Delta = t_f/\kappa$. Fig.~\ref{fig: consolidated_scatter_1-4_0.1} shows the results for controllers with $\Delta = t_f/\kappa \approx 0.1$ ($\Delta \in [0.094, 0.117]$). Fig.~\ref{fig: consolidated_scatter_1-4_0.2} shows $\mathcal{B}_\text{vu}$ versus $\varepsilon$ for controllers with $t_f/\kappa \in [0.195,0.234]$. In both cases the larger system (the five-qubit chain of Problem~$4$) yields controllers that exhibit both higher fidelity error and greater differential sensitivity as measured by $\mathcal{B}_\text{vu}$. At the other end of the spectrum, Fig.~\ref{fig: consolidated_scatter_1-4_0.1} shows the controllers optimized for Problem~$1$ (the less-complex two-qubit system) clustering in the lower-left, the region for the lowest error and differential sensitivity bounds. Though the basic trend of controllers for larger systems showing a less desirable performance-robustness profile is seen in both figures, the difference is less pronounced for the larger values of $\Delta$. Interestingly, in Fig.~\ref{fig: consolidated_scatter_1-4_0.1} we observe a cluster of Problem~$4$ controllers with similar profiles to those of Problem~$2$, and in Fig.~\ref{fig: consolidated_scatter_1-4_0.2} we see the same similarity for Problem~$3$ and Problem~$4$ controllers. This shows that while a more complex (larger) system generally results in lower fidelity and higher sensitivity controllers for a similar control architecture, it is not the only determining factor. Rather a combination of factors, e.g., $t_f$ and $\kappa$, also affect controller robustness, which can be leveraged to achieve the best results.

\begin{figure*}
\subfloat[\label{fig: prob6_vs_prob9_top}]{\includegraphics[width=0.49\textwidth]{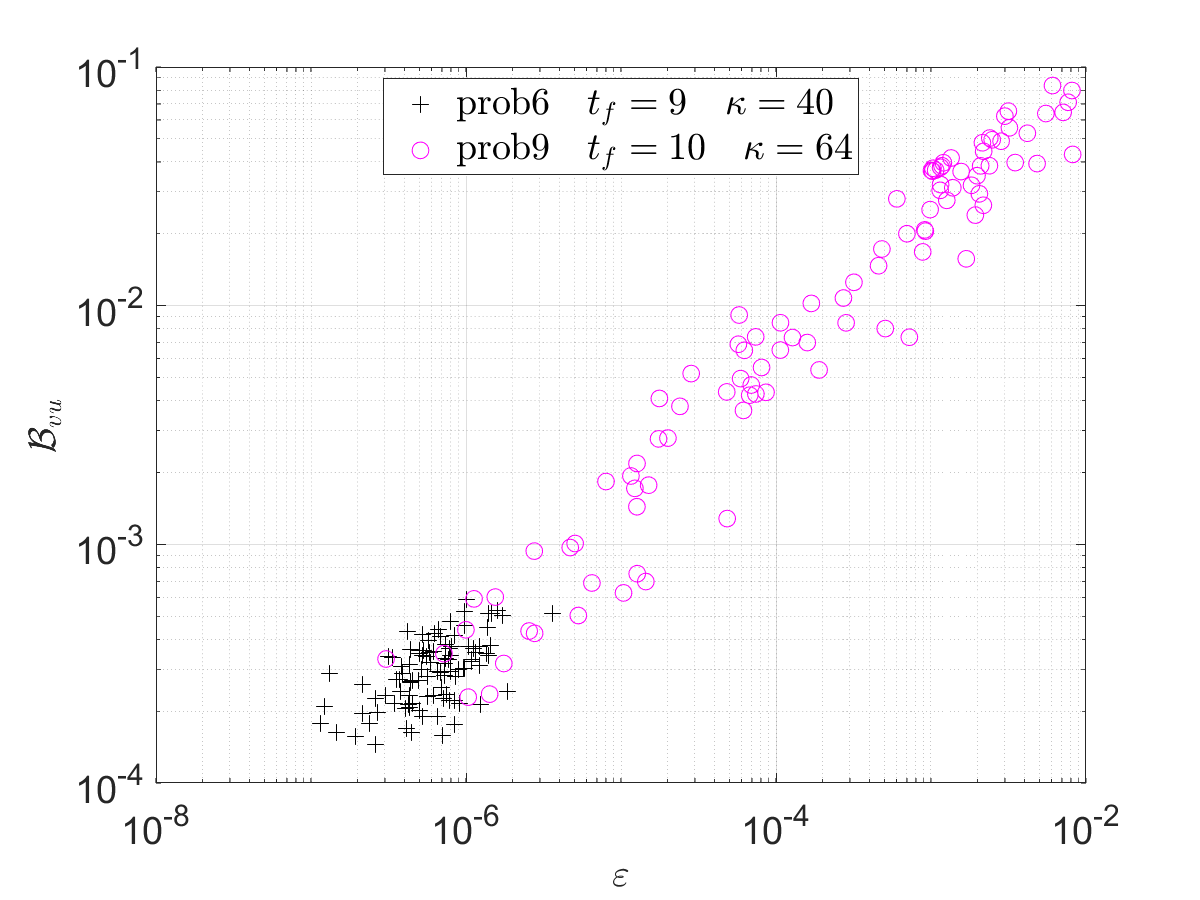}} \hfill
\subfloat[\label{fig: prob6_vs_prob9_bottom}]{\includegraphics[width=0.49\textwidth]{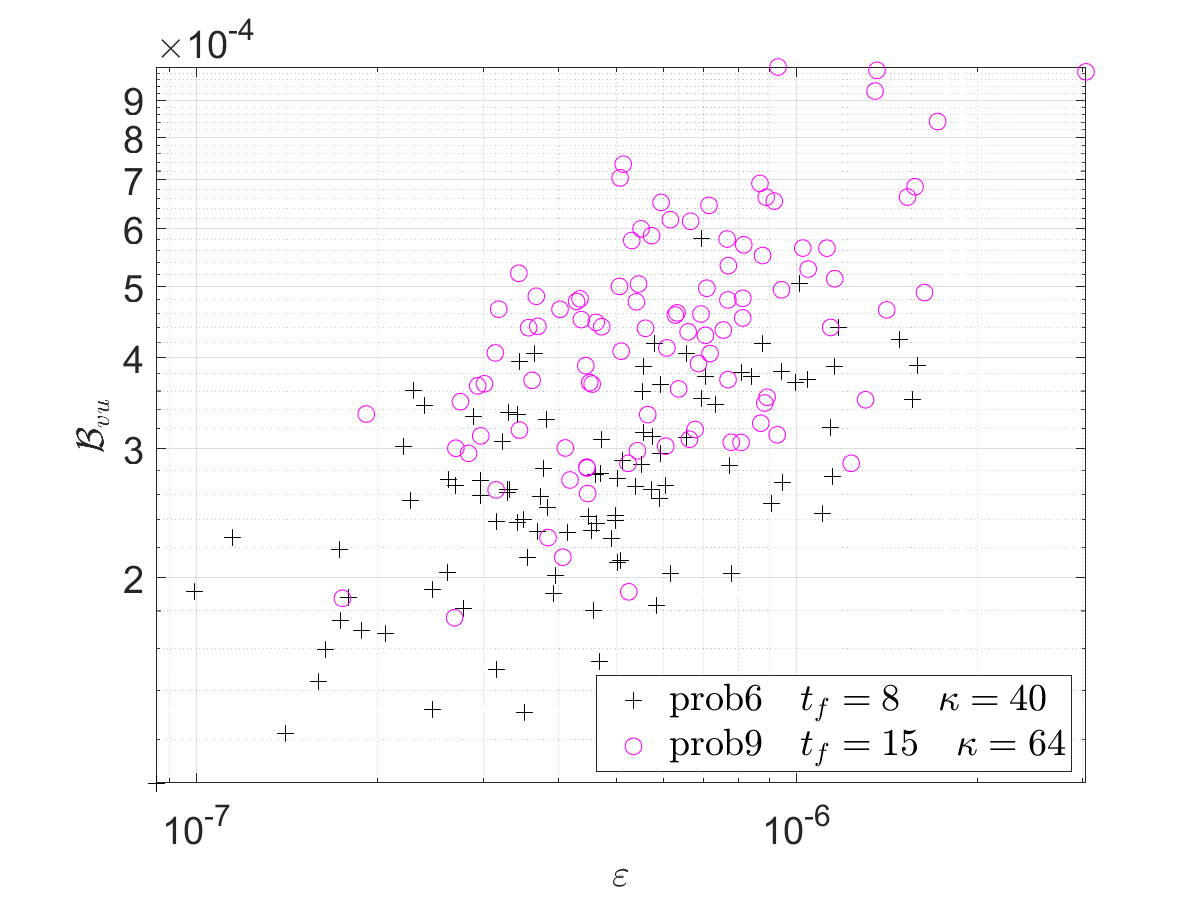}}
\caption{Plot of $\mathcal{B}_\text{vu}$ versus $\varepsilon$ on a log-log scale for controllers for Problem~$6$ and~$9$ grouped by similar values of $t_f/\kappa$, where (a) shows $t_f/\kappa<0.1750$ and (b) shows $t_f/\kappa>0.200$. The plot in (a) shows that individual spin addressability yields controllers with a superior robustness-performance profile than those for control on only the first spin for $t_f/\kappa$ between $0.1562$ and $0.1750$. However, the plot in (b) shows that for $\kappa$ in the range of $0.200$ to $0.2344$, both control implementations yield similar performance-robustness profiles.} \label{fig: prob6_vs_prob9}
\end{figure*}

\begin{figure*}
\subfloat[\label{fig: prob5_vs_prob8_top}]{\includegraphics[width=0.49\textwidth]{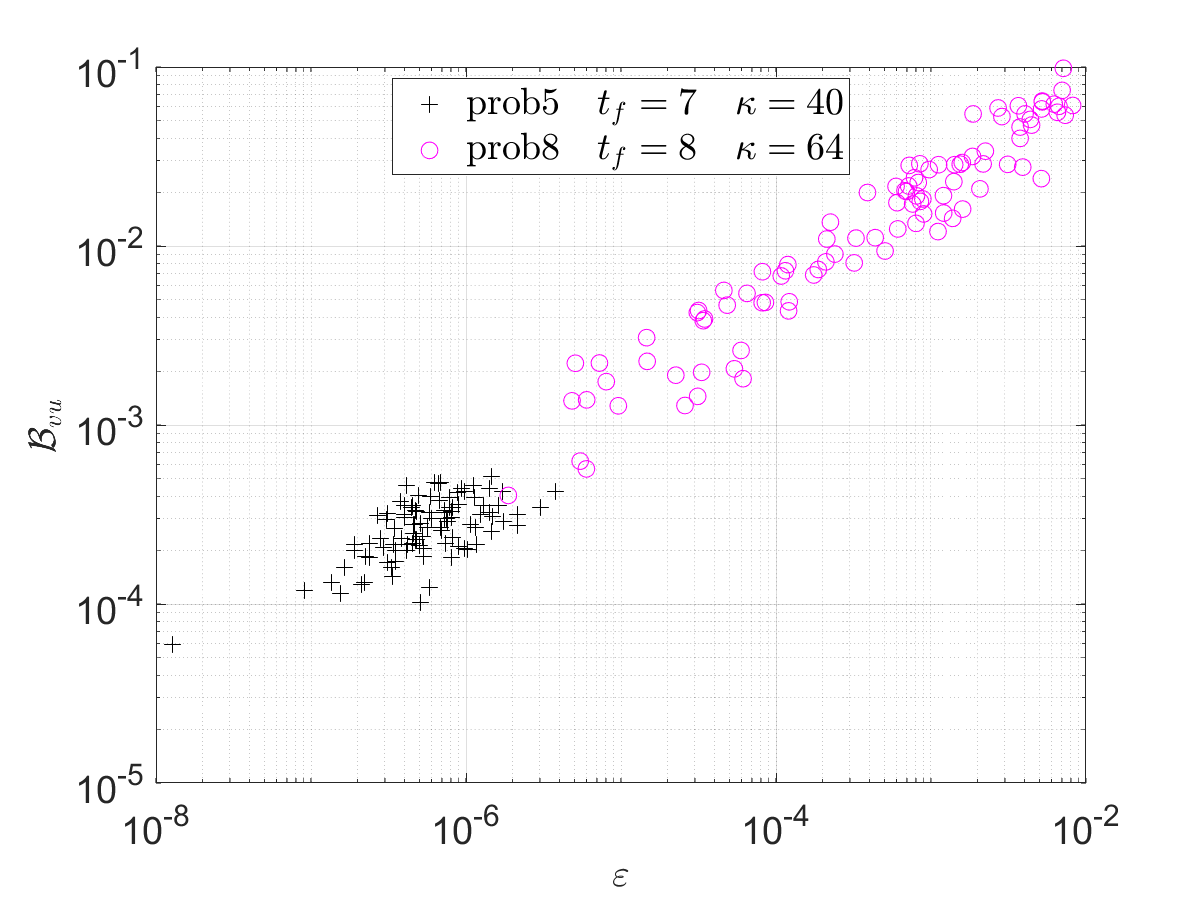}}
\hfill
\subfloat[\label{fig: prob5_vs_prob8_bottom}]{\includegraphics[width = 0.49\textwidth]{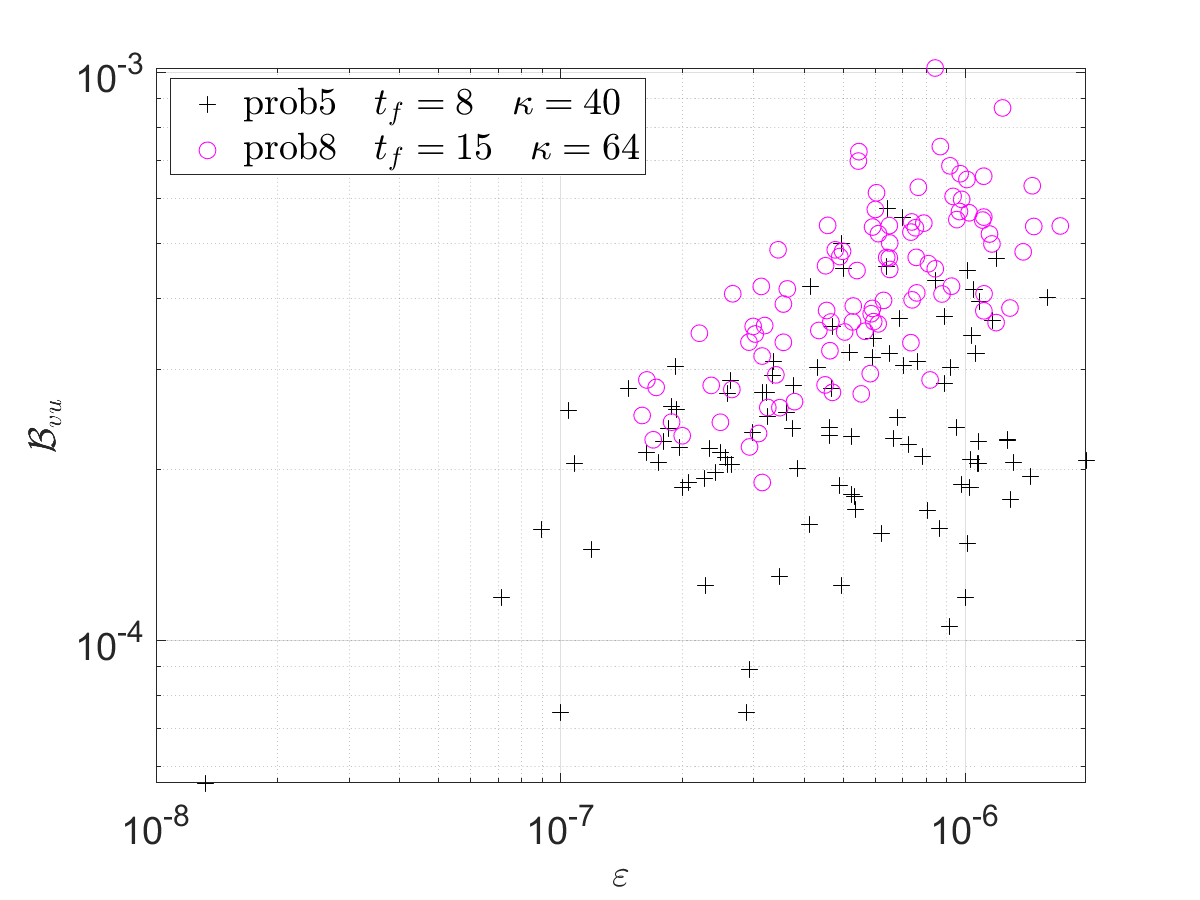}}
\caption{Plot of $\mathcal{B}_\text{vu}$ versus $\varepsilon$ on a log-log scale for controllers from Problems~$5$ and~$8$ for (a) $t_f/\kappa<0.1750$ and (b) $t_f/\kappa>0.200$. In (a), we see that the implementation of individual qubit control yields controllers with a superior robustness-performance profile than those with control on only the first qubit for $t_f/\kappa$ between $0.1562$ and $0.1750$. However, (b) shows that for $\kappa$ in the range of $0.200$ to $0.2344$, both control implementations yield similar performance, with only slightly smaller differential sensitivity for the individual spin addressable case.}\label{fig: prob5_vs_prob8}
\end{figure*}

\begin{figure*}[t]
\subfloat[\label{fig: prob4_vs_prob7}]
{\includegraphics[width = 0.32\columnwidth]{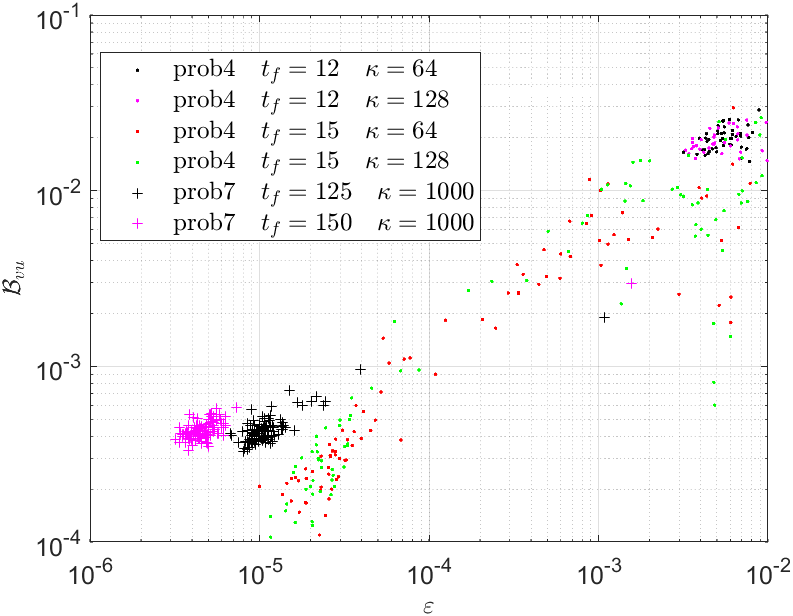}}
\hfill
\subfloat[\label{fig: prob7}]{\includegraphics[width = 0.32\columnwidth]{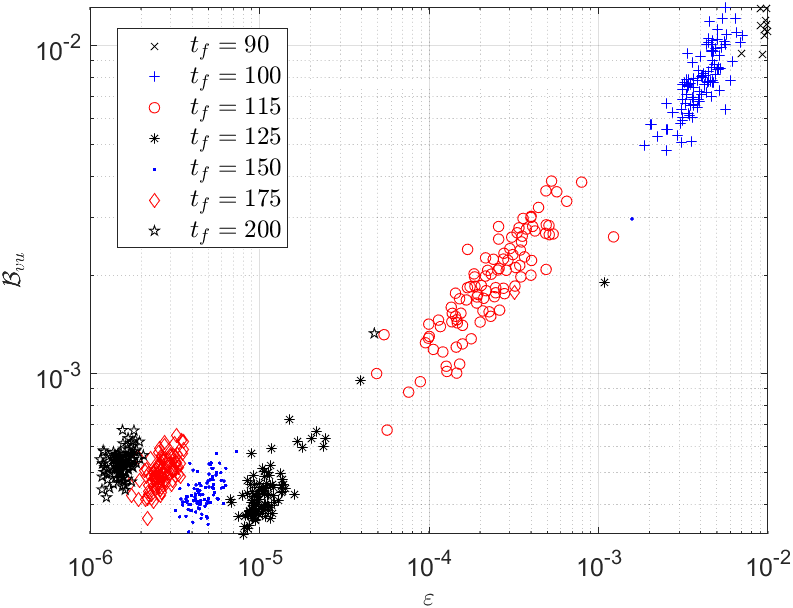}}
\hfill
\subfloat[\label{fig: prob4}]{\includegraphics[width = 0.32\columnwidth]{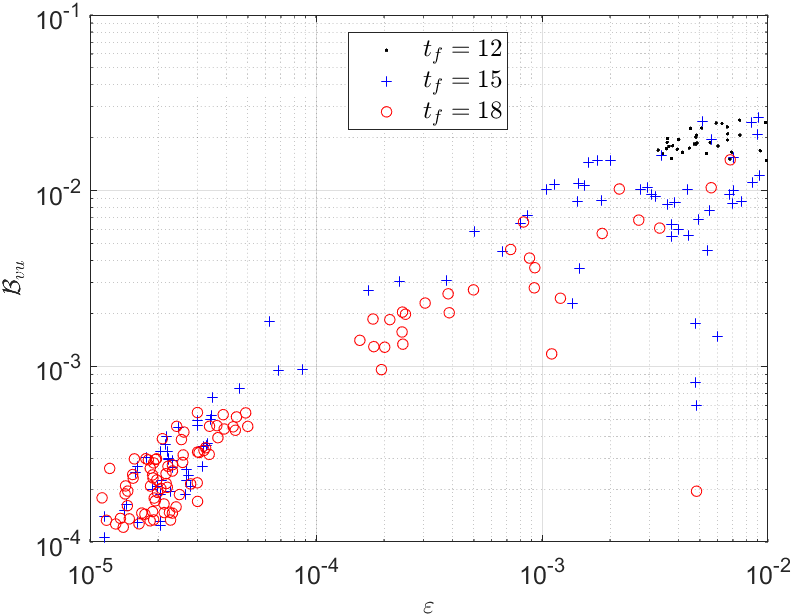}}
\caption{Comparison of $\mathcal{B}_\text{vu}$ versus $\varepsilon(t_f)$ for the Ising ZZ $5$-Chains of Problem 4 and Problem 7. (a) Controllers for five-qubit gate Problems~$4$ and~$7$ for a range of $t_f$ and $\kappa$ combinations. (b) Problem~$7$ controllers for $\kappa = 1000$ and increasing $t_f$. (c) Problem~$4$ controllers for fixed $\kappa = 128$ and increasing $t_f$. While both problems have the same target QFT gate, Problem 4 implements individual spin addressable control, and Problem 7 implements simultaneous control on all spins.}
\end{figure*}

\begin{figure*}[t]
\subfloat[\label{fig: gate_comp_top}]{\includegraphics[width=0.49\textwidth]{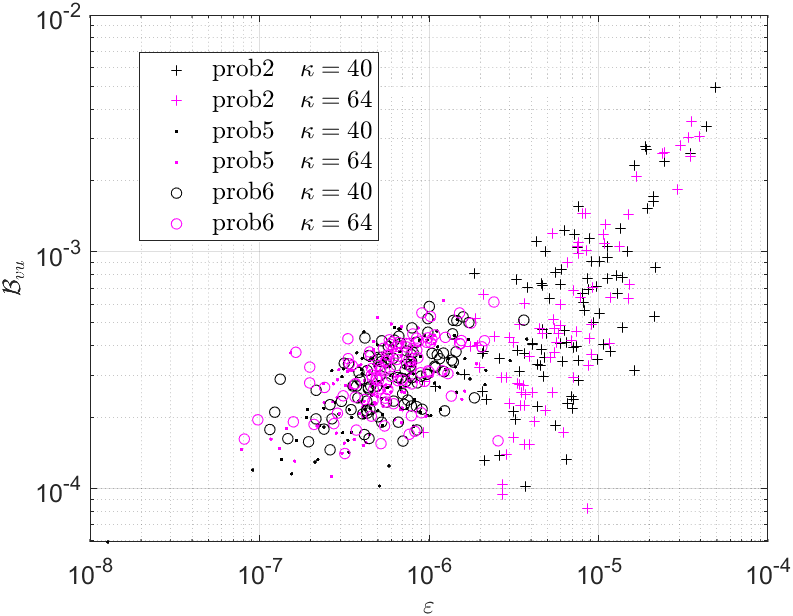}} \hfill
\subfloat[\label{fig: gate_comp_bottom}]{\includegraphics[width = 0.49\textwidth]{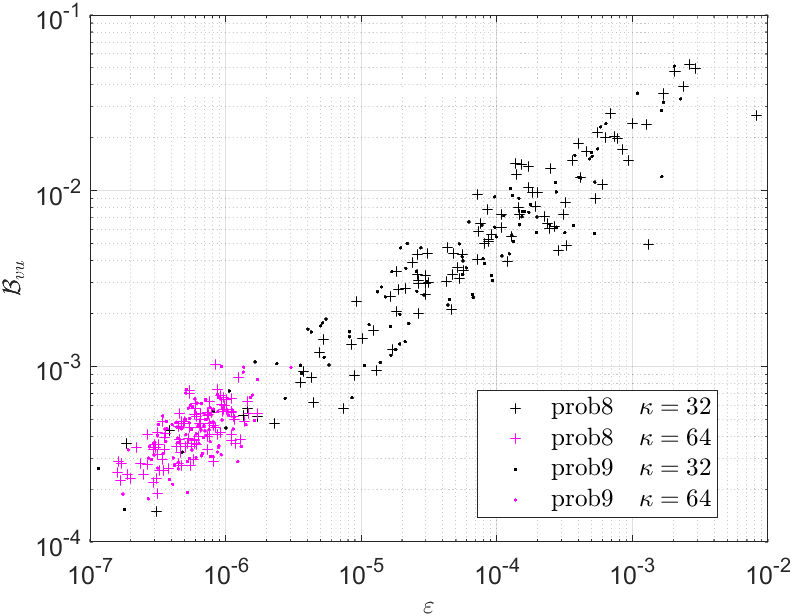}}
\caption{Plot of $\mathcal{B}_\text{vu}$ versus $\varepsilon$ for three-chain controllers grouped by similar control implementation. (a) Plot of $\mathcal{B}_\text{vu}$ versus $\varepsilon$ on a log-log scale for the controllers in Problems~$2$,~$5$, and~$6$ with $t_f = 7$. (b) Plot of $\mathcal{B}_\text{vu}$ versus $\varepsilon$ on a log-log scale for the controllers in Problems~$8$ and~$9$ with $t_f = 15$. The plot in (a) shows that for individual spin-addressability, the Ising-coupled system results in greater fidelity error than the Heisenberg coupled system. Additionally, despite the difference in target gate between Problem~$5$ and Problem~$6$, both yield controllers with comparable performance-robustness profiles. The plot in (b) shows the results for Problems~$8$ and~$9$ where the $t_f/\kappa$ ratio impacts the performance and robustness more than the difference in target gate.} \label{fig: gate_comp}
\end{figure*}

\subsection{Effect of Control Type and Target Gate}

We examine the impact of different types of control implementation on the robustness for systems with the same number of qubits. Fig.~\ref{fig: prob5_vs_prob8} shows a comparison of the controllers for Problems~$5$ and~$8$.   Both systems have three qubits with nearest-neighbor Heisenberg coupling and the target is a QFT gate, but in Problem~$5$ each qubit can be addressed individually while in Problem~$8$ direct control is restricted to the first qubit. Likewise, Fig.~\ref{fig: prob6_vs_prob9} shows a comparison with the controllers for Problems~$6$ and~$9$, again two three-qubit systems with nearest-neighbor Heisenberg coupling with the same randomly selected target unitary gate but differing in spin addressability. Given the reduced degrees of freedom in the control for Problems~$8$ and~$9$, we expect that controllers with good fidelity and sensitivity characteristics should be more difficult to produce. Fig.~\ref{fig: prob5_vs_prob8_top} shows that this holds for a value of $t_f/\kappa < 0.175$. However, for $t_f/\kappa > 0.200$ Fig.~\ref{fig: prob5_vs_prob8_bottom} shows that the resulting fidelities for the controllers are very similar, but the most robust controllers are associated with the individual spin-addressable implementation. The same analysis of the controllers for Problems~$6$ and~$9$ shows similar results for the random unitary target gate (see Figs~\ref{fig: prob6_vs_prob9_top} and~\ref{fig: prob6_vs_prob9_bottom}).

This suggests that with the appropriate $t_f$ and discretization of the control pulses the control based on a single addressable spin is just as effective as the paradigm where all spins can be controlled. However, beyond a specific lower limit on $t_f/\kappa$, finding controllers with similar performance and robustness characteristics becomes less tenable. The main reason for this is that the effects of a control acting on one qubit at end of the register are propagated along the linear register by the fixed coupling between adjacent qubits. This imposes lower bounds on the amount of time required to achieve controllability, the value of which mainly depends on the size of the system and the strength of the interactions. While the increased gate operation times may be disadvantageous, this approach has benefits related to computational resource requirements by reducing the number of controls to optimize over. Moreover, there are potential improvements in experimental feasibility, as controlling a single qubit can be technologically easier than controlling every qubit in a register. Furthermore, single-qubit control can provide real-world robustness improvements by reducing the number of control channels and sources of uncertainty associated with their use.

Fig.~\ref{fig: prob4_vs_prob7} shows robustness versus fidelity error comparison of controllers for Problems~$4$ and~$7$. While both systems are five-qubit registers with Ising coupling between adjacent qubits, in Problem~$4$ each qubit is individually controllable while in Problem~$7$ the control fields act globally, i.e., the controls affect simultaneous rotations on all qubits but a linear detuning is applied to the qubit register. Though the control implementation for Problem~$7$ might seem less effective due to the reduced degrees of freedom, as Fig.~\ref{fig: prob4_vs_prob7} shows, this is not necessarily the case.  While for Problem~$4$ controllers with $t_f=15$ exhibit the lowest bounds $\mathcal{B}_\text{vu}$ on the differential sensitivity (see Fig.~\ref{fig: prob4}), the Problem~$7$ controllers (see Fig.~\ref{fig: prob7}) have similar or better fidelity, with error on the order of $10^{-5}$ or less, for only slighter greater differential sensitivity. As with the previous case, this suggests that optimizing for a control paradigm with reduced degrees of freedom has the potential to yield controllers that exhibit good performance and low sensitivity with the same benefits in reduction of computational overhead and increased robustness in implementation.

We also examine the effect of differing target gates on systems of the same size with the same controller implementation. Fig.~\ref{fig: gate_comp} shows the plots of $\mathcal{B}_\text{vu}$ versus $\varepsilon$ for the three-qubit problems, grouped by control implementation. Fig.~\ref{fig: gate_comp_top} shows that for the QFT target gate. Controllers for the Heisenberg-coupled system of Problem~$5$ perform better than controllers for the Ising-coupled system of Problem~$2$. However, the smallest bounds on the differential sensitivity are achieved for the Ising system. A comparison of the plots for Problems~$5$ and~$6$ shows that the effect of the target gate on the resulting robustness and performance of the controllers is negligible with both controllers clustering in the same area of the plot. Fig.~\ref{fig: gate_comp_bottom} shows the same for limited control implementation of Problems~$8$ and~$9$. The same trend is obtained in this plot, so, at least for the problems considered here, the effect of the target gate has no noticeable impact on the performance-robustness characteristics. Rather the $t_f/\kappa$ ratio has a much more noticeable impact on the production of controllers with desirable properties versus those on the other end of the spectrum.

\section{Log-Sensitivity and Fundamental Limitations}\label{sec: fundamental_limitations}

\begin{figure*}
\subfloat[\label{fig: log_sens_classical}]{\includegraphics[width=0.49\textwidth]{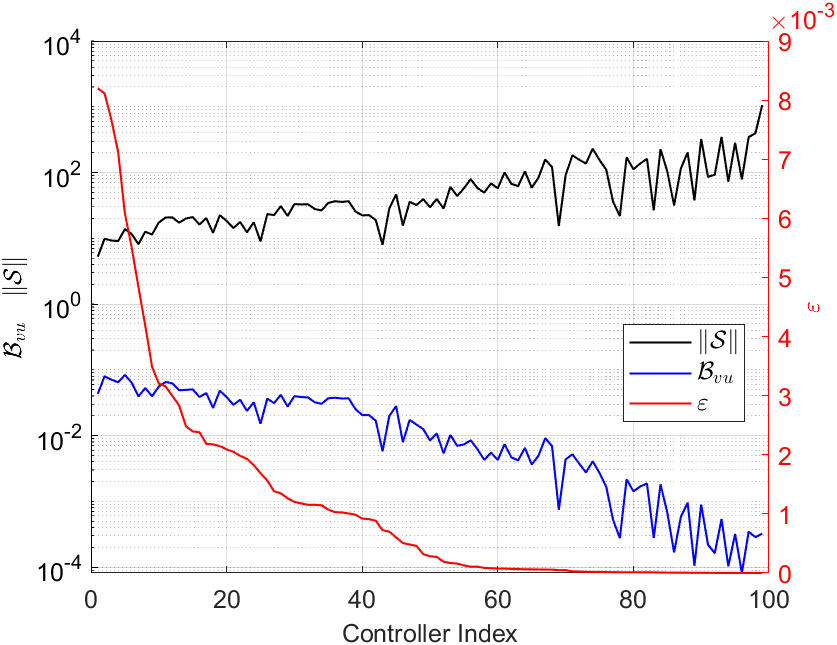}}
\hfill
\subfloat[\label{fig: log_sens_not_so_classical}]{\includegraphics[width = 0.49\textwidth]{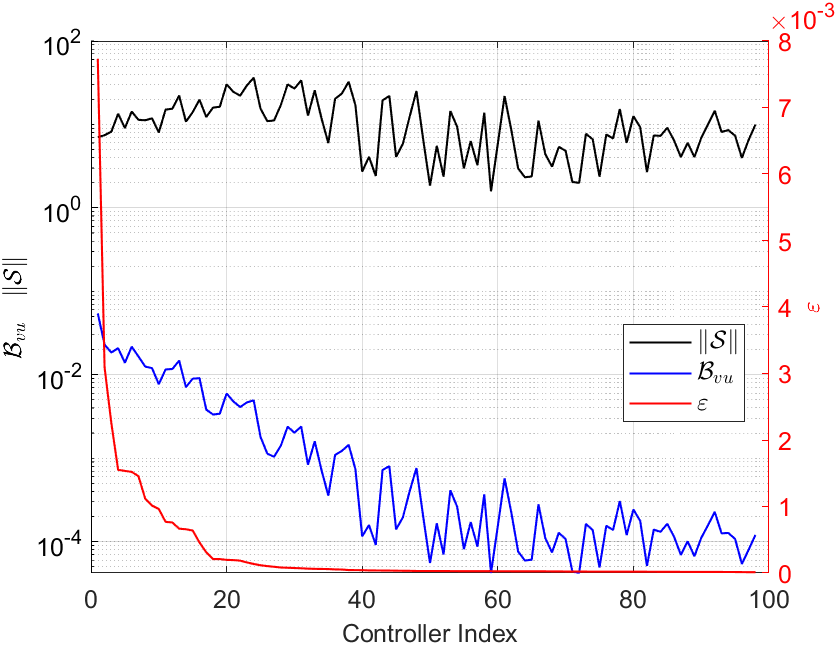}}
\caption{Plot of log-sensitivity, sensitivity bounds, and nominal fidelity error versus controller index. (a) Plot of $\|\mathcal{S}\|$, $\mathcal{B}_\text{vu}$, and $\varepsilon$ versus controller index on a semi-log scale for the three-chain Problem~$9$, $t_f=10$, $\kappa=64$. (b) Plot of $\|\mathcal{S}\|$, $\mathcal{B}_\text{vu}$, and $\varepsilon$ versus controller index on a semi-log scale for the four-chain Problem~$3$, $t_f=12$, $\kappa=40$. The plot in (a) shows the trade-off between log-sensitivity and fidelity error expected from classical control, while (b) shows one of the two exceptions in the data set.}
\end{figure*}

The previous sections have shown that there is generally no trade-off between the differential sensitivity bounds and the performance as measured by the fidelity error. For sake of comparison with classical control limitations, we analyze the trade-off (or lack thereof) between performance (as measured by $\varepsilon$) and robustness (as measured by the log-sensitivity), proceeding along the same lines as that done in Refs.~\onlinecite{data_set_1,data_set_2}.  The log-sensitivity is essentially the differential sensitivity of the \emph{logarithm} of the error,
which equates to the differential sensitivity of the error divided by the error.  Note that the definitions in Refs.~\onlinecite{data_set_1} and \onlinecite{data_set_2} differ slightly.  The former definition aligns more closely with classical control theory but fails for perturbations around a parameter with nominal value $0$.  To circumvent this problem, the definition of the log-sensitivity for a perturbation structured as $H_\mu$ was amended in Ref.~\onlinecite{data_set_2}.  This is the definition used here:
\begin{equation}\label{eq: log-sens}
  \mathcal{S}_{\mu} =
  \left. \frac{\partial \ln(\tilde{\varepsilon}_{\mu})}{\partial \delta} \right|_{\delta = 0} =  \frac{1}{\varepsilon} \zeta_{\mu}.
\end{equation}
\textcolor{black}{Now let the unit vector in the parameter space $\mathbf{s}_\mu$ only take values in $\set{e_k}_{m=0}^M$ where ${e_k}$ is a natural basis vector (i.e. $\mathbf{s}_1 = (1,\hdots,0)^T$ through $\mathbf{s}_M = (0,\cdots,1)^T$). Then $\mathcal{S}_\mu$ is the log-sensitivity of the fidelity error to uncertainty in one of the principal directions $\set{0,1, \hdots, M}$. To make the analysis tractable,} we take the $2$-norm of the vector of $M+1$ possible log-sensitivity values for a given controller as $\|\mathcal{S}\| = \sqrt{\sum_{\mu = 0}^M \mathcal{S}^2_\mu}$.
This provides a single value to use as measure of the log-sensitivity for a given controller. We then test the level of concordance between the error and $\|\mathcal{S}\|$ based on the Kendall $\tau$ rank correlation coefficient. Specifically, we execute a one-tailed test for anti-concordance with a null hypothesis $\mathcal{H}_0$ of no correlation between $\varepsilon$ and $\|\mathcal{S}\|$ and an alternative hypothesis $\mathcal{H}_-$ of negative rank correlation. For the opposite conclusion, we test for concordance in the same manner, but with an alternative hypothesis $\mathcal{H}_+$ indicated by positive rank correlation. For both tests we set the significance level at $95\%$ ($p = 0.05$).

Of the $86$ total test cases, $11$ fail to meet the significance threshold. Of the $75$ remaining tests, all but two reject $\mathcal{H}_0$ in favor of $\mathcal{H}_-$. The two cases that reject $\mathcal{H}_0$ in favor of $\mathcal{H}_+$ and indicate a non-classical trend are the Problem~$3$ and $t_f=12$ cases. Fig.~\ref{fig: log_sens_classical} shows the trend between the log-sensitivity, as measured by $\|\mathcal{S}\|$, and the differential sensitivity, as measured by the bound $\mathcal{B}_\text{vu}$, with the fidelity error for Problem~$9$, $t_f=10$, and $\kappa = 64$. The negative trend between the log-sensitivity $\|\mathcal{S}\|$ and fidelity error $\varepsilon$ is readily apparent and borne out the by the Kendall $\tau$ of $-0.693$. 
Conversely, Fig.~\ref{fig: log_sens_not_so_classical} shows one of the two non-classical cases for Problem~$3$, $t_f=12$, and $\kappa=40$. The positive correlation between $\|\mathcal{S}\|$ and $\varepsilon$ is \textcolor{black}{not readily apparent visually but is borne out by a Kendall $\tau$ of $0.270$.}

In summary, despite the positive trend between the differential sensitivity and $\varepsilon(t_f)$, the expected classical limitations between the performance and robustness, as measured by the log-sensitivity observed in earlier work~\cite{data_set_1},\cite{data_set_2} still hold in the main. Future work should consider which metric, the pure differential sensitivity or the normalized log-sensitivity, is a more useful measure of robustness in the context of quantum control.

\section{Conclusion}\label{sec:conc}

The differential sensitivity of the error and recently derived upper bounds on the differential sensitivity for time-domain control with piecewise-constant functions were introduced as a measure of robustness for quantum control implementation. The results were applied to understand and quantify the effect of uncertainty in the system and control Hamiltonian on the performance of controllers optimized for a variety of dynamic quantum gate implementation problems. The data revealed an unexpected concordance between the upper bounds of the differential sensitivity of the fidelity error and the fidelity error for a broad range of systems, optimization targets, and controllers, which suggests that if the upper bound on the differential sensitivity of the error is used as a measure of robustness then there is no trade-off between robustness and performance, i.e., the best-performing controllers are also the most robust. This suggests that it is not necessary to explicitly optimize for robustness in these cases. Given the prevalence of explicitly optimizing for robustness in the literature, this is a surprising result.

Comparison of robustness versus error plots for a large number of controllers for different systems and optimization targets indicate that both the performance and robustness of controllers decreases with increasing system size. Specifically, five-qubit gates result in larger fidelity error and differential sensitivity bounds than smaller qubit systems, although in most cases the controllers continued to achieve both high performance and robustness. Moreover, the analysis suggests there are generally no significant robustness and performance reductions when control is restricted to limited local control of a single qubit or global control without local addressing compared to full local control of all qubits. This is promising as system architectures with limited control tend to be technologically easier to realize.

There are limitations to the applicability of robustness measures based on the differential sensitivity.  It is likely to be a useful measure to assess and compare robustness of controllers in the high-fidelity, small perturbation regime, such as the implementation of high-fidelity quantum gates subject to small uncertainties.  For problem involving large perturbations and larger errors, other robustness-infidelity measures may be useful to assess robustness post-synthesis~\cite{RIM}.  More work is necessary to compare different robustness measures to establish when they should be used and under which conditions optimization for robustness and fidelity is beneficial. 

Furthermore, we only considered structured perturbations to the system or control Hamiltonians, which are a significant source of error in many applications, but there are other sources of error, which have not been considered here.  Some, such as decoherence, can be modeled using structured perturbations~\cite{data_set_2}.  Other errors such as those arising from imperfect realization of piecewise-constant controls, e.g., due to rise and settling times or bandwidth limitations may be less amenable to treatment using structured perturbations and should be considered independently using complementary methods, e.g., to compensate for nonlinear distortions in the controls due to hardware limitations~\cite{PhysRevApplied.19.064067}.

\section*{Data Availability}
All controller data, robustness analysis code, and results data associated with this manuscript is available at \url{https://qyber.black/spinnet/code-differential-sensitivity-bounds-for-dynamic-control}.  

\section*{Acknowledgements}

We acknowledge the support of the Supercomputing Wales project (SCW2103), which is part-funded by the European Regional Development Fund (ERDF) via the Welsh Government. \textcolor{black}{C.A.W., F.C.L. and S.M.S. acknowledge funding from EPSRC via grant number EP/Y004728/1.}

\section*{Author Declarations}

The authors have no conflicts to disclose.

\section*{Appendix: Fundamental Limits in Classical Control}

The positive correlation between the upper bounds on the differential sensitivity and the nominal fidelity error $\varepsilon(t_f)$ calls into question whether the systems under consideration circumvent the fundamental limitations on performance and sensitivity established by classical control~\cite{Safonov_1981}.  Specifically, in a classical frequency domain context, there is a trade-off between the tracking error and the normalized logarithmic sensitivity of the tracking error, in the form of the identity $S(s) + T(s) = I$, where $S(s)$ is proportional to the tracking error and $T(s)$ is a measure of the normalized logarithmic-sensitivity (Eq.~\eqref{eq: log-sens}) of the closed loop system to parameter variation~\cite{Dorf2011}.  For a purely sinusoidal input, $s = i\omega$ and $S(i \omega)$ and $T(i \omega)$ are the frequency-dependent gains from the reference input to the error signal and system output, respectively, obtained via the Laplace transform.

\section*{References}

\bibliographystyle{apsrev.bst}
\bibliography{refs}

\end{document}